\documentclass{article}
\usepackage{epsfig} 
\usepackage{amsmath}
\usepackage{amsfonts}
\usepackage{graphicx}

\usepackage{a4}

\usepackage{amsmath}

\begin{document}

\title{A non-Abelian Chern-Simons--Yang-Mills-Higgs system \\
in $3+1$ dimensions}
\author{{\large Francisco Navarro-L\'erida}$^{\ddagger}$,
{\large Eugen Radu}$^{\diamond}$
and {\large D. H. Tchrakian}$^{\star \dagger}$ 
\\ 
$^{\ddagger}${\small Dept.de F\'isica At\'omica, Molecular y Nuclear, Ciencias F\'isicas,}
\\
{\small Universidad Complutense de Madrid, E-28040 Madrid, Spain}
\\  
$^{\diamond}${\small Departamento de F\'\i sica da Universidade de Aveiro and I3N,}
\\
{\small Campus de Santiago, 3810-183 Aveiro, Portugal}
 \\ 
$^{\star}${\small School of Theoretical Physics, Dublin Institute for Advanced Studies,}\\
{\small 10 Burlington Road, Dublin 4, Ireland }\\
{$^\dagger$\small Department of Computer Science, NUI Maynooth, Maynooth, Ireland}}

\date{\today}
\newcommand{\dd}{\mbox{d}}
\newcommand{\tr}{\mbox{tr}}
\newcommand{\la}{\lambda}
\newcommand{\ka}{\kappa}
\newcommand{\f}{\phi}
\newcommand{\F}{\Phi}
\newcommand{\al}{\alpha}
\newcommand{\ga}{\gamma}
\newcommand{\Ga}{\Gamma}
\newcommand{\de}{\delta}
\newcommand{\si}{\sigma}
\newcommand{\Si}{\Sigma}
\newcommand{\bnabla}{\mbox{\boldmath $\nabla$}}
\newcommand{\bomega}{\mbox{\boldmath $\omega$}}
\newcommand{\bOmega}{\mbox{\boldmath $\Omega$}}
\newcommand{\bsi}{\mbox{\boldmath $\sigma$}}
\newcommand{\bchi}{\mbox{\boldmath $\chi$}}
\newcommand{\bal}{\mbox{\boldmath $\alpha$}}
\newcommand{\bpsi}{\mbox{\boldmath $\psi$}}
\newcommand{\brho}{\mbox{\boldmath $\varrho$}}
\newcommand{\beps}{\mbox{\boldmath $\varepsilon$}}
\newcommand{\bxi}{\mbox{\boldmath $\xi$}}
\newcommand{\bbeta}{\mbox{\boldmath $\beta$}}
\newcommand{\ee}{\end{equation}}
\newcommand{\eea}{\end{eqnarray}}
\newcommand{\be}{\begin{equation}}
\newcommand{\bea}{\begin{eqnarray}}

\newcommand{\ii}{\mbox{i}}
\newcommand{\e}{\mbox{e}}
\newcommand{\pa}{\partial}
\newcommand{\Om}{\Omega}
\newcommand{\vep}{\varepsilon}
\newcommand{\bfph}{{\bf \phi}}
\newcommand{\lm}{\lambda}
\def\theequation{\arabic{equation}}
\renewcommand{\thefootnote}{\fnsymbol{footnote}}
\newcommand{\re}[1]{(\ref{#1})}
\newcommand{\R}{{\rm I \hspace{-0.52ex} R}}
\newcommand{\N}{{\sf N\hspace*{-1.0ex}\rule{0.15ex}%
{1.3ex}\hspace*{1.0ex}}}
\newcommand{\Q}{{\sf Q\hspace*{-1.1ex}\rule{0.15ex}%
{1.5ex}\hspace*{1.1ex}}}
\newcommand{\C}{{\sf C\hspace*{-0.9ex}\rule{0.15ex}%
{1.3ex}\hspace*{0.9ex}}}
\newcommand{\eins}{1\hspace{-0.56ex}{\rm I}}
\renewcommand{\thefootnote}{\arabic{footnote}}

\maketitle


\bigskip

\begin{abstract}
We study spherically symmetric solutions of an 
$SO(5)$ Chern-Simons--Yang-Mills-Higgs system in $3+1$ dimensions.
The Chern-Simons densities  are defined
in terms of both Yang-Mills fields and a $5$-component isomultiplet Higgs.
The $SO(3)\times{SO(2)}$ solutions are analysed in a systematic way, by employing numerical methods.
These finite energy configurations possess both 
electric and magnetic global charges,  
differing radically, however, from Julia-Zee dyons.
When two or more of these Chern-Simons densities are present in the
Lagrangian, solutions with vanishing electric charge but nonvanishing electrostatic potential may exist.
\end{abstract}
\medskip
\medskip

\section{Introduction}
The 'usual' Chern-Simons (CS) densities are defined in all odd dimensions~\cite{Jackiw:1985}, both Euclidean or Minkowskian.
This is because their definition relies on that of
the Chern-Pontryagin (CP) density in one dimension higher, which is an even dimension. Recently however,
'new' Chern-Simons(--Higgs)~\cite{Tchrakian:2010ar,Radu:2011zy} densities in all odd and even
dimensions have been proposed. The aim of the present work is to employ such Chern-Simons--Higgs (CSH) terms to construct
solitons.

As a first application of these new CSH terms, we carry out this task in ($3+1$)-dimensional Minkowski spacetime. This choice
offers a novel example of the use of a Chern-Simons density in $even$ spacetime dimensions, and secondly, $3+1$ is the most relevant
physical dimension. In addition to the magnetic the magnetic flux,
the presence of a Chern-Simons terms results (as usual) in an electric flux.

Ever since the work of \cite{Deser:1982vy} on topologically massive $SU(2)$ Yang--Mills (YM) theory in $2+1$ dimensions, systems in $2p+1$ dimensions
featuring a Chern-Simons term have been studied. Some of these are ($2+1$)-dimensional Higgs models \cite{Hong:1990yh,Jackiw:1990aw,NavarroLerida:2008uj}
supporting Abelian and non-Abelian vortices, while others, like \cite{Brihaye:2010wp}, turn out to be truncations of gauged supergravities
\cite{Cvetic:2000nc}.

Before stating the definitions of the Chern-Simons--Higgs (CSH) densities in $3+1$ dimensions, to be employed in the present work, we review the
general definition of these new Chern-Simons densities for the sake of being self-contained here.
The definition of a Chern-Simons density $\Omega_{\rm CS}$ in a ($2n+1$)-dimensional spacetime is extracted from the ($2n+2$)-dimensional
Chern-Pontryagin densitiy $\Omega_{\rm{CP}}$, which is by construction a total divergence 
\be
\Omega_{\rm{CP}}[F]={\bnabla}\cdot\bOmega[A,F] \, , \label{CP_density}
\ee
$A$ and $F$ being the Yang-Mills connection and curvature, respectively. 

The CS density is then defined as the $(2n+2)-$th component $\Omega_{2n+2}$, of the $(2n+2)$-component density $\bOmega$,
\be
\Omega_{\rm CS}[A,F]\stackrel{\rm def}=\Omega_{2n+2}[A,F] \, , \label{CS_density_odd}
\ee
in one dimension lower, namely, in $2n+1$ dimensions. These CS densities,  are $gauge\ variant$.

Since the CP densities $\Omega_{\rm{CP}}$ are defined only in $even$, $2n+2$ dimensions, it follows that the corresponding CS densities
$\Omega_{\rm{CS}}$ are defined only in $odd$, ($2n+1$)-dimensional spacetimes. 

To define Chern-Simons densities in $even$ dimensional spacetimes, it would be natural to extract these from the analogues of the Chern-Pontryagin
densities defined in $odd$ dimensions, which are also total divergence. These new CP densities $\hat\Omega_{\rm{CP}}$, are by
construction~\cite{Tchrakian:2010ar,Radu:2011zy} also total divergence
\be
\hat\Omega_{\rm{CP}}={\bnabla}\cdot\hat\bOmega \, . \label{td}
\ee
Such densities can be constructed by subjecting the CP density in some higher $even$
dimension, to dimensional descent to some (lower) residual dimension, say $D+2$, which can be
odd~\footnote{This descent does not have to be down to an odd residual dimension, $D+2$, as is the case in the present work.
The residual dimension $D+2$ can just as well be even.}.

The reduced CP densities $\hat\Omega_{\rm{CP}}$ in $D+2$ dimensions, which are reviewed in \cite{Tchrakian:2010ar}, in addition to the curvature $F$,
depend also on a Higgs field $\F$ resulting from the breakdown of symmetry in the dimensional descent, as well as on its covariant derivative $D\F$.
Like the CP density in the higher dimensions, they are also $gauge\ invariant$. Their most remarkable property is that
the density $\hat\bOmega$ in Eq.~\re{td} is gauge $covariant$ in $odd$, and gauge $variant$ in $even$, (residual) dimensions,
\bea
\hat\Omega_{\rm{CP}}[F,D\F,\F^2]&=&{\bnabla}\cdot\hat\bOmega[F,D\F,\F^2]\ ,\quad {\rm for\ odd}\ \ D+2 \, , \label{tdo} \\
\hat\Omega_{\rm{CP}}[F,D\F,\F^2]&=&{\bnabla}\cdot\hat\bOmega[A,F,D\F,\F^2]\ ,\quad {\rm for\ even}\ \ D+2\,.\label{tde}
\eea
The definitions of the new Chern-Simons densities now follow naturally, as the $(D+2)$-th components of $\hat\bOmega$ in Eqs.~\re{tdo} and \re{tde},
respectively
\bea
&&\hat\Omega_{\rm CS}\stackrel{\rm def}=\hat\Omega_{D+2}[F,D\F,\F^2]\ ,\quad {\rm for\ even}\ \ D+1\, , \label{hatcso}\\
&&\hat\Omega_{\rm CS}\stackrel{\rm def}=\hat\Omega_{D+2}[A,F,D\F,\F^2]\ ,\quad {\rm for\ odd}\ \ D+1 \, ,\label{hatcse}
\eea
in a $(D+1)$-dimensional spacetime.

In the present work, our attention is restricted to the case of $D=3$, namely, to the case of four-dimensional Minkowski spacetime, where we will construct
static, spherically symmetric, solitons of a Yang-Mills--Higgs system featuring (new) Chern-Simons terms,
which carry both electric and magnetic global charges. The new CS terms being those extracted from the
dimensionally descended (from some higher even dimensions) Chern-Pontryagin density in residual $3+2(=D+2)$ dimensions, the gauge group and the multiplicity
of the Higgs field are fixed. This is explained in detail in Ref. \cite{Tchrakian:2010ar}.

In this preliminary work,
the multiplet structure in the $3+1$ dimensional model is chosen to be the most economical one consistent with the possibility of constructing soliton
solutions in these dimensions, and, such that the ``new Chern-Simons'' density is nonvanishing. In the present work the descent resulting in the CS density in $3+1$ dimensions
starts from the $n$-th Chern-Pontryagin density in $2n$ dimensions and ends in $D+2=5$ dimensions, for $n=3,4$ and $D=3$. Following Ref. \cite{Tchrakian:2010ar}, the (residual)
YM field $F_{\mu\nu}=(F_{ij},F_{i0})$ and the Higgs field $\F$ are both $4\times 4$ antihermitian matrices which we choose to take their values in the chiral Dirac representation of $SO(6)$.

This most economical choice is
\bea
F_{\mu\nu}&=&F_{\mu\nu}^{\al\beta}\,\Si_{\al\beta} \,, \label{so5-6}\\
\F&=&\f^{\al}\,\Si_{\al 6}\ ,\quad \al=i,4,5;\ i=1,2,3\,,\label{5vec}
\eea
$(\Si_{\al\beta},\Si_{\al 6})$ being the $4\times 4$ chiral representation matrices of $SO(6)$.
The spin matrices $\Sigma_{\mu\nu}=(\Sigma_{\al\beta},\Sigma_{\al 6})$ used here are
\be
\label{sigma1}
\Sigma_{\mu\nu}=-\frac14\Sigma_{[\mu}\,\tilde\Sigma_{\nu]}\,,
\ee
with $$\Sigma_{i}=-\tilde\Sigma_{i}=i\gamma_i\ ,\
\Sigma_{4}=-\tilde\Sigma_{4}=i\gamma_4\ ,\ \Sigma_{5}=-\tilde\Sigma_{5}=i\gamma_5\ ,\
\Sigma_{6}=+\tilde\Sigma_{6}=\eins,$$ defined in terms of the usual Dirac gamma
matrices ($\gamma_{i},\ga_4,\ga_5$), $i=1,2,3$.

There is a tower of dynamical CS densities on $(3+1)$-dimensional Minkowski spacetime that one can employ. Each
of these is arrived at $via$ the dimensional reduction of the $n-$th Chern-Pontryagin density on $K^5\times S^{2n-5}$,
$n\ge 3$. The CS density on Minkowski $M^4$ is then defined as the $5-$th component of the density on the residual space $K^5$,
in one dimension lower. Here, we display the first two members of this tower, pertaining to $n=3$ and $n=4$, respectively,
\bea
\Omega_{\rm CS}^{(1)}&=&i\,\epsilon^{\mu\nu\rho\si}\mbox{Tr}\,\F\,F_{\mu\nu}\,F_{\rho\si}\label{CS1} \,,\\
\Omega_{\rm CS}^{(2)}&=&i\,\epsilon^{\mu\nu\rho\si}\mbox{Tr}\bigg[
\F\left(\eta^2\,F_{\mu\nu}F_{\rho\si}+\frac29\,\F^2\,F_{\mu\nu}F_{\rho\si}+\frac19\,F_{\mu\nu}\F^2F_{\rho\si}\right)
\nonumber\\
&&\qquad\qquad\qquad\qquad
-\frac29\left(\F D_{\mu}\F D_{\nu}\F-D_{\mu}\F\F D_{\nu}\F+D_{\mu}\F D_{\nu}\F\F\right)F_{\rho\si}\bigg]\, , \label{CS2}
\eea
where $\epsilon^{\mu\nu\rho\si}$ is the Levi-Civita tensor in Minkowski spacetime.

Inasfar as Eqs.~\re{CS1} and \re{CS2} present CP violating dynamics, the (Higgs) scalar may be interpreted as an axion~\cite{Peccei:1977ur,Peccei:1977hh} like scalar.

\section{The models and field equations}
The models considered will feature the Chern-Simons terms $\Omega_{\rm CS}^{(1)}$ and $\Omega_{\rm CS}^{(2)}$, Eqs.~\re{CS1}-\re{CS2},
augmented by the Yang-Mills-Higgs (YMH) sector.
As explained in the Introduction of this preliminary work, we will choose the YMH sector to be the usual one consisting of the traces of squares
of the $SO(5)$ YM curvature, Eq.~\re{so5-6}, and the covariant derivative of the Higgs field, Eq.~\re{5vec}, plus the usual quartic Higgs potential.

The Lagrangian densities we will consider are
\be
\label{CSYMH}
{\cal L}
= {\cal L}_{\rm YMH} +\ka_1\,\Omega_{\rm CS}^{(1)}+\ka_2\,\Omega_{\rm CS}^{(2)}\,,
\ee
where $\Omega_{\rm CS}^{(1)}$ and $\Omega_{\rm CS}^{(2)}$ are given by Eq.~\re{CS1} and Eq.~\re{CS2}, respectively, and,
\be
\label{CSYMH1}
{\cal L}_{\rm YMH}=\mbox{Tr}\left[\frac14F_{\mu\nu}^2-\frac12\,D_{\mu}\F^2-\frac{\la}{2}\,(\F^2+\eta^2\eins)^2\right]\,,
\ee
where $D_\mu = \partial_\mu + [A_\mu, \cdot]$.
Here $\ka_1$ and $\ka_2$ represent the corresponding CS coupling constants, $\la$ is the Higgs potential coupling constant, and $\eta$ denotes the vacuum
expectation value of the Higgs field.

We will seek solutions with both magnetic and electric global charges for the systems Eq.~\re{CSYMH}, to which we will loosely refer as dyons in the following.
But what we have proposed here is quite different from the Julia-Zee (JZ) dyon~\cite{Julia:1975ff}. In the latter, the electric component $A_0$
of the gauge connection and the Higgs field $\F$ both take their values in the algebra of the
same gauge group, $SO(3)$, while here $A_0$ and $\F$ have
entirely different multiplet structures as implied in Eqs.~\re{so5-6}-\re{5vec}. It should be emphasised that the main difference is not that here we have
the gauge group $SO(5)$ instead of $SO(3)$ of the JZ dyon, but rather that the electric component of the gauge connection results from the
Chern-Simons dynamics exploited in \cite{Hong:1990yh,Jackiw:1990aw}.

The equations of motion resulting from the variations of the Lagrangian with respect to the YM potential and the Higgs field are
\bea
D_{\mu}F^{\mu\nu}+[\F,D^{\nu}\F]&=&2\,i\,\ka_1\,\vep^{\mu\nu\rho\si}\,\{F_{\rho\si},D_{\mu}\F\}\,,\label{varYM}\\
D_{\mu}D^{\mu}\F-\la\{\F,(\F^2+\eta^2\eins)\}&=&i\,\ka_1\,\vep^{\mu\nu\rho\si}\,F_{\mu\nu}\,F_{\rho\si}\,,\label{varH}
\eea
respectively. $\{\ , \ \}$ denotes the anticommutator. These equations, Eqs.~\re{varYM} and \re{varH}, are written only for the Lagrangian with
$\kappa_2=0$ in Eq.~\re{CSYMH}. This is because the expressions for the right-hand sides of the corresponding equations for $\kappa_2\neq 0$ are very
cumbersome.

It is clear from the Gauss-Law equation, namely for the $\nu=0$ component of Eq.~\re{varYM}, that when the Chern-Simons coupling constants vanishes,
so will the component $A_0$ of the gauge connection, resulting in a vanishing electric charge. This is a typical feature of Chern-Simons-Higgs
dyons~\cite{Hong:1990yh,Jackiw:1990aw,NavarroLerida:2008uj}.

The electric field here, $E_i\stackrel{\rm def.}=F_{i0}$ is in general a non-Abelian quantity, leading to the definition of the flux.
This definition is equivalent (up to a sign) to the general definition of electric charge for non-Abelian fields \cite{Corichi:1999nw,Corichi:2000dm} computed as  
\be
\label{elec}
\displaystyle Q^{YM} = \frac{1}{4\pi}   {\oint}_{S^\infty} \bigg[-\sum_{i=1}^3 \mbox{Tr}(F^2_{i0})  \bigg]^{1/2} \, dS~,
\ee

Likewise, our definition of the non-Abelian magnetic charge is given by
\be
\label{mag}
\displaystyle P^{YM} = \frac{1}{4\pi} \oint_{S^\infty} \bigg[-\sum_{i=1}^3 \mbox{Tr}(\tilde{F}^2_{i0}) \bigg]^{1/2} \, dS~,
\ee
where $\tilde F$ is the Hodge dual of the gauge field. 

It should be emphasised here that the magnetic charge Eq.~\re{mag} is a global charge, and not a topological charge. The reason is that our Higgs multiplet
here is a $5$-component isovector, rather than the $3$-component isovector in the case of the usual t'Hooft-Polyakov monopole. Indeed, unlike the $monopole$,
of the Georgi-Glashow model in which the gauge field breaks down to an Abelian field due to the symmetry breaking mechanism, here, our definition of the magnetic
flux in Eq.~\re{mag} does not involve the Higgs field.

The monopole charge of the Georgi-Glashow model is
\be
\label{tH-P}
\mu=\frac{1}{4\pi}\,\vep_{ijk} \int_{S^\infty}\mbox{Tr}\,\F\,F_{ij}\, dS_k\,,
\ee
which presents a lower bound on the energy integral. This is not the case with the solutions in this work. Nothwistanding, one might refer to these solutions
loosely as dyons, understanding that in the soliton literature the word dyon normally implies topological stability for the magnetic sector.



\section{Imposition of spherical symmetry and one dimensional subsystems}

\subsection{The general case}
The
static spherically symmetric Ansatz for the Higgs field $\F$ and the YM connection
$A_{\mu}=(A_0,A_i)$ reads
\bea
\F&=&2\eta\left(\f^M\,\Si_{M6}+\f^6\,\hat x_j\,\Si_{j6}\right)\,,
\label{higgs}
\\
A_0&=&-(\vep\chi)^M\,\hat x_j\,\Sigma_{jM}-
\chi^{6}\,\Sigma_{45}\,, \label{a0p}\\
A_i&=&\left(\frac{\xi^{6}+1}{r}\right)\Sigma_{ij}\hat x_j+
\left[\left(\frac{\xi^M}{r}\right)\left(\delta_{ij}-\hat x_i\hat x_j\right)+
(\vep A_r)^M\,\hat x_i\hat x_j\right]\Sigma_{jM}+\nonumber\\
&&\qquad\qquad\qquad\qquad\qquad\qquad\qquad\qquad +A_r^{6}\,
\hat x_i\,\Sigma_{45}\,, \label{aip}
\eea
in which $i,j=1,2,3$ and $M=4,5$.
We can label the functions $(\f^M,\f^{6})\equiv\vec\f$,
$(\chi^M,\chi^{6})\equiv\vec\chi$, $(\xi^M,\xi^{6})\equiv\vec\xi$ and $(A_r^M,A_r^{6})\equiv\vec A_r$
like four isotriplets $\vec\f$, $\vec\chi$, $\vec\xi$ and $\vec A_r$, all depending on
the $3$-dimensional spacelike radial variable $r$. $\vep$ is the two
dimensional Levi-Civita symbol.

The parametrization used in the Ansatz, Eqs.~\re{higgs}-\re{aip}, results in a gauge
covariant expression for the YM curvature $F_{\mu\nu}=(F_{ij},F_{i0})$ and the gauge covariant derivative of the Higgs $D_{\mu}\F=(D_{i}\F,D_{0}\F)$
\bea
\nonumber
F_{ij}&=&\frac{1}{r^2}\left(|\vec\xi|^2-1\right)\Sigma_{ij}+
\frac1r\left[D_r\xi^{6}+\frac1r\left(|\vec\xi|^2-1\right)\right]
\hat x_{[i}\Sigma_{j]k}\hat x_{k}+
\frac1rD_r\xi^M\hat x_{[i}\Sigma_{j]M}\,,
\\
F_{i0}&=&-\frac1r\,\xi^M(\vep\chi)^M\,\Sigma_{ij}\hat x_j+\frac1r\,
\left[\xi^{6}(\vep\chi)^M-\chi^{6}(\vep\xi)^M\right]\Sigma_{iM}
\nonumber
\\
&-&\left\{(\vep D_r\chi)^M+\frac1r\,
\left[\xi^{6}(\vep\chi)^M-\chi^{6}(\vep\f)^M\right]\right\}
\hat x_i\hat x_j\Sigma_{jM}
-D_r\chi^{6}\,\hat x_i\,\Sigma_{45}\,,
\label{fi0}
\\
\nonumber
(2\eta)^{-1}D_i\F&=&-\frac1r(\vec\xi\cdot\vec\f)(\delta_{ij}-\hat{x_i}\hat{x_j})\,\Si_{j6}
+D_r\f^M\hat{x_i}\,\Si_{M6}+D_r\f^6\,\hat{x_i}\hat{x_j}\,\Si_{j6}\,,
\label{DiF}
\\
\nonumber
(2\eta)^{-1}D_0\F&=&\f^M(\vep\chi)^M\hat x_j\,\Sigma_{j6}-\left[\f^{6}(\vep\chi)^M
-\chi^{6}(\vep\f)^M\right]\Sigma_{M6}\,,
\label{D0F}
\eea
in which we have used the notation
\be
\nonumber
\label{covp}
D_r\xi^a=\pa_r\xi^a+\vep^{abc}\,A_r^b\,\xi^c\quad,\quad
D_r\chi^a=\pa_r\chi^a+\vep^{abc}\,A_r^b\,\chi^c\quad,\quad
D_r\f^a=\pa_r\f^a+\vep^{abc}\,A_r^b\f^c \,,
\ee
as the $SO(3)$ covariant derivatives of the three triplets
$\vec\xi$, $\vec\chi$ and $\vec\f$ with respect to the one dimensional, and hence trivial,
$SO(3)$ gauge connection $\vec A_r$.

Substituting Eq.~\re{higgs} and Eqs.~\re{fi0} in the CS densities, Eqs.~\re{CS1}-\re{CS2},
the resulting reduced one dimensional CS Lagrangian for the first CS term, Eq.~\re{CS1}, is
\be
\label{redCS1}
\omega_{\rm{CS}}^{(1)}=-8\ka_1\,\eta\,\left[(|\vec\xi|^2-1)\,\vec\f\cdot D_r\vec\chi-
2(\vec\xi\times\vec\chi)\cdot(\vec\f\times D_r\vec\xi)\right]\,,
\ee
and that for the second CS term, Eq.~\re{CS2}, is
\bea
\nonumber
\omega^{(2)}&=&-\frac{16}{3}\ka_2\eta^3\bigg(\frac{|\vec\f|^2-3}{2}\left[(|\vec\xi|^2-1)\,\vec\f\cdot D_r\vec\chi-
2(\vec\xi\times\vec\chi)\cdot(\vec\f\times D_r\vec\xi)\right]
-\bigg[
(|\vec\xi|^2-1)(\vec\f\times{\vec\chi})\cdot(\vec\f\times{D_r\vec\f})
\\
\label{redCS2}
&&-
(\vec\xi\cdot\vec\f)^2\,\vec\f\cdot{D_r\vec\chi}
+2(\vec\xi\cdot\vec\f)\left((\vec\f\times{\vec\chi})\cdot(\vec\f\times{D_r\vec\xi})+
(\vec\xi\times{\vec\chi})\cdot(\vec\f\times{D_r\vec\f})
\right)\bigg]\bigg) \,,
\eea
where $\omega^{(i)}_{CS} = \kappa_i r^2 \Omega^{(i)}_{CS}$.

For completness, we give also the expression of the corresponding expression 
of the reduced YMH static Lagrangian ($L_{\rm{YMH}}=r^2{\cal L}_{\rm{YMH}}$, Eq.~\re{CSYMH1}):
 \bea
\label{lagp=1}
L_{\rm{YMH}}&=&-\frac12\left(2\,|D_r\vec\xi|^2+
\frac{1}{r^2}\left(|\vec\xi|^2-1\right)^2\right)
+\frac12\left(r^2\,|D_r\vec\chi|^2
+2\,|(\vec\xi\times\vec\chi)|^2\right)\nonumber\\
&&\qquad+2\,\eta^2\,r^2\,\left(|(\vec\f\times\vec\chi)|^2-
\left[|D_r\vec\f|^2+\frac{2}{r^2}(\vec\xi\cdot\vec\f)^2\right]\right)-2\la\,\eta^4\,(1-|\vec\f|^2)^2\,,
\eea
the first line pertaining to the YM fields and the second to the Higgs.

The variation with respect to the trivial $SO(3)$ gauge connection $\vec A_r$ does not give rise to an equation of motion, but rather gives
$constraint\ equations$. Furthermore, the $SO(3)$ freedom in this Ansatz results in an invariance at the fixed point of the $2$-sphere, due to
which only two of the components of each of the three triplets $(\vec\xi,\vec\chi,\vec\f)$ are independent functions. We thus
end up with $6$ equations of motion for the functions of $r$,
\be
\label{trunc}
\vec\xi=( \tilde w,0,w)\ ,\ \vec\chi=( \tilde V,0,V)\ ,\ \vec\f=(\tilde h,0,h) \,,
\ee
in addition to the constraint equations.

\subsection{The $SO(3)\times{SO(2)}$ case}

To simplify the picture, in what follows, we shall construct only those solutions for which 
\begin{eqnarray}
\label{truncation}
\tilde w=\tilde V=\tilde h=0 \ ,
\end{eqnarray}
$i.e.$, our solutions describe the $SO(3)\times{SO(2)}$
submultiplet of the $SO(5)$ Yang-Mills field. These solutions will possess both electric and magnetic fields. 
Moreover, they will describe $finite$ $energy$ solutions, by virtue of the chosen asymptotic values of the fields, consistent with analyticity.
With these asymptotic values, also the electric and magnetic fluxes Eqs.~\re{elec} and \re{mag}, are nonvanishing.

The electric field leads in general to a non-Abelian flux.
In the restricted case we are considering ($\tilde w=\tilde V=\tilde h=0$), the electric field is proportional to
$\Sigma_{45}$ , so it is the $SO(2)$ submultiplet of the $SO(5)$ field,
Eq.~\re{fi0}, namely, the quantity 
$\pa_r\chi^6=V'$ 
appearing in front of the algebra basis $\Si_{45}$. This leads us to a natural definition of the
electric charge $Q$ of the solution as 
\be
Q = \lim_{r \to \infty} r^2\frac{d V}{d r} \, . 
\label{Q_computation} 
\ee 
This definition, which is equivalent (up to a sign) to the general definition of electric charge for non-Abelian fields \cite{Corichi:1999nw,Corichi:2000dm}, and is computed
from Eq.~\re{elec} as  
\be
\label{elec1}
\displaystyle Q^{YM} = 
\lim_{r \to \infty}  r^2 \sqrt{\left|\frac{d\vec\chi}{dr}\right|^2 + \frac{2}{r^2}|\vec\xi \times \vec \chi|^2}  \, , 
\ee
which for our restricted case reduces to $Q^{YM} = |Q|$.

Likewise, the (scalar) magnetic fluxin  the restricted case we are considering is computed from Eq.~\re{mag} as
\be
\label{mag1}
\displaystyle P^{YM} = 
\lim_{r \to \infty} \sqrt{\left(1-\left|\vec\xi\right|^2\right)^2 + 2 r^2 \left| \frac{d \vec\xi}{dr}\right|^2}  \, .
\ee
For this restricted case and taking into account the asymptotic behaviour of the solutions
(see next section), one can see that the magnetic charge is $P^{YM}=1$.

We emphasised at the end of the previous section that the (scalar) magnetic charge is a global charge, and not the topological charge \re{tH-P}
of the t'Hooft-Polyakov $magnetic$ $monopole$. Indeed, one can readily evaluate the flux integral Eq.~\re{tH-P} for the spherically symmetric fields Eqs.~\re{higgs},
\re{a0p}, and \re{aip} (and not only for the restricted case  $\tilde w=\tilde V=\tilde h=0$), 
which turns out to vanish. Our magnetic charge is not a topological charge.

Let us close this section
by mentioning that within this $SO(3)\times{SO(2)}$ truncation,
the system with the first CS term, $\Omega_{\rm CS}^{(1)}$,
is effectively described by a YMH-Maxwell system,
\begin{eqnarray}
\label{truncation1}
 {\cal L} =\mbox{Tr}\left[\frac14 \mathcal{F}_{\mu\nu}^2-\frac12\,D_{\mu}  \F^2-\frac{\la}{2}\,(\F^2+\eta^2\eins)^2\right]
 -\frac14  {f}_{\mu\nu}^2+i\kappa_1\epsilon^{\mu\nu\rho\sigma}f_{\mu\nu} \mbox{Tr}~\F \mathcal{F}_{\rho\sigma} \, ,
\end{eqnarray}
where this time $\mathcal{F}_{\mu\nu}$ and $\F$ are $SO(3)$ fields. 
The $SO(3)$ and $U(1)$ fields interact only via the CS term,
the $U(1)$ field being purely electric, with $f_{r0}=V'$.
In this theory, the monopole charge Eq.~\re{tH-P} does not vanish, such that the solutions 
are topologically stable.
No similar effective model could be constructed for the case of the second  CS term, $\Omega_{\rm CS}^{(2)}$.

\section{Restricted field equations and boundary conditions}

\subsection{The equations and an effective model}

Substituting the Ansatz, Eqs.~\re{higgs}-\re{aip}, in the field equations, Eqs.~\re{varYM}-\re{varH}, (with $\kappa_2$ included and together with our
gauge choice Eq.~\re{trunc} and the restriction $\tilde w=\tilde V=\tilde h=0$) 
the following $3$ equations are obtained~\footnote{The constant $\eta$ may be set to any
nonvanishing value by rescaling the radial coordinate $r$. In what follows we have chosen it to be $1/2$.}:
\bea
&& \frac{d^2V}{dr^2}  +\frac{2}{r} \frac{dV}{dr}
 +\frac{2}{r^2} w  h  \left[ 4\kappa_1 + \kappa_2 (1-h^2)\right] \frac {dw}{dr} 
-\frac{1}{r^2}  \left[  \left( 4\kappa_1  +\kappa_2   ( 1- h ^2) \right) ( 1-w^2) 
+2\kappa_2 w^2 h^2 \right] \frac {dh}{dr} = 0 \,, 
\label{eq_b}  
\\
&&\frac{d^2w}{dr^2}  + w 
 \left[h \left( 4\kappa_1  +\kappa_2   ( 1- h ^2) \right) \frac {d V}{dr}  + 
\frac {1- w^2}{r^2} -  h^2 \right] = 0 \,,
\label{eq_a} 
\\
&&\frac{d^2h}{dr^2} + \frac{2}{r} \frac {dh}{dr}
 + \frac{ \lambda}{2} h (1-h^2) -\frac{2 w^2 h}{r^2}
 - \frac{1}{r^2}  \left[  \left( 4 \kappa_1 + \kappa_2 (1-h^2)  \right)  ( 1-
 w^2) +2  \kappa_2
  w^2 h^2 \right] \frac {dV}{dr} = 0 \,. \label{eq_h}
\eea

However, Eq.~(\ref{eq_b}) has a total derivative structure, which implies the existence 
of the first integral\footnote{Here ar arbitrary
integration constant is set to zero as required by the finite energy conditions.}
\begin{eqnarray}
\label{firstintegral}
\frac{dV}{dr}=\frac{h}{r}
\left [
4\kappa_1(1-w^2)+\kappa_2
\left(
(1-w^2)+h^2(w^2-\frac{1}{3})
\right) 
\right].
\end{eqnarray}
After replacing the above relation in Eqs.~(\ref{eq_a} ), (\ref{eq_h} ),
we find that the system is effectively described by the reduced Lagrangian density 
\begin{eqnarray}
\label{Leff}
{\cal L}_{eff}={\cal L}_{YMH}^{(0)}+{\cal L}_{YMH}^{(int)} \, ,
\end{eqnarray}
with
\begin{eqnarray}
\label{Leff1}
&&
{\cal L}_{YMH}^{(0)}=w'^2+\frac{(1-w^2)^2}{2r^2}+w^2 h^2+
\frac{1}{2}r^2 h'^2+\frac{1}{8}r^2 \lambda (1-h^2)^2,
\\
&&
{\cal L}_{YMH}^{(int)}=
\left[
\kappa_1\frac{2\sqrt{2}h(1-w^2)}{r}
+\kappa_2
\frac{h}{\sqrt{2}r}
\left(
(1-\frac{h^2}{3})(1-w^2)+\frac{2}{3}h^2w^2
\right)
\right ]^2~,
\end{eqnarray}
which corresponds to  $SO(3)$ magnetic monopoles with an extra-interaction term as given by ${\cal L}_{YMH}^{(int)}$.

In order to obtain regular dyonic solutions we impose an appropriate set of boundary conditions. At the origin the
functions and their derivatives must satisfy
\be
w|_{r=0} = -1 \, , \, \left. ~~\frac{dV}{dr}\right|_{r=0} = 0 \, , \,~~ h|_{r=0}=0 \, , \label{bc_origin}
\ee
while their asymptotic values are
\be
w|_{r=\infty} = 0 \, , \,~~ \left. V\right|_{r=\infty} = 0 \, , \,~~ h|_{r=\infty}=1 \, . \label{bc_infty}
\ee
The second condition in Eq.~(\ref{bc_infty}) fixes the gauge freedom
for the electric potential.
 Under these conditions the energy of the solutions\footnote{Note that the CS term does not contribute to the energy density
 of the solutions.}
\bea
E &=& \int_0^\infty \left[ \left(\frac{dw}{dr}\right)^2 + \frac{1}{2} \frac{(1-w^2)^2}{r^2} +
\frac{1}{2} r^2  \left(\frac{dV}{dr}\right)^2 \right.
 \left. + \frac{1}{2}  r^2  \left(\frac{dh}{dr}\right)^2 +  w^2h^2 + \frac{\lambda}{8} r^2 (1-h^2)^2  \right] dr \, , \label{energy}
\eea
is finite.

As implied by Eq.~(\ref{firstintegral}), 
the electric field of the solutions is
induced by the CS term, the magnetic monopoles acquiring an electric charge 
 \be
Q = \frac{2 \kappa_2}{3} + 4 \kappa_1 \, .
\label{Q_expression}
\ee

Remarkably, since the signs of $\kappa_1$ and $\kappa_2$ are free, one may have solutions with a vanishing electric charge $Q$ but with a nonvanishing electric
component of the gauge potential $V(r) \neq 0$, 
$i.e.$, a nonvanishing electric field.
 Note that $V(r)$ is identically zero only if both $\ka_1$ and $\ka_2$ are
zero simultaneously.

\subsection{Asymptotic analysis}
Unfortunately, the system with $V(r)\neq 0$ does not seem to possess exact solutions\footnote{A promising direction appeared
to 
be to construct them as a perturbation around the BPS
monopoles $(\lambda=0)$, by treating $\kappa_i$ as small parameters.
However, the final $linear$ equations could not be solved even in this case.}.
However, approximate expressions can be written both as $r\to 0$ and as $r\to \infty$.
The expansions at the origin of the solutions look rather simple
\bea
w(r)= -1 + a_2 r^2 + O(r^4) ~,~~
V(r)= b_0 + \frac{1}{3}h_1 (12 \kappa_1 a_2 + 3 \kappa_2 a_2 + \kappa_2 h_1^2) r^2 + O(r^4),~~~
h(r)= h_1 r + O(r^3),
\eea
where $a_2$, $b_0$, and $h_1$ are constants. 

The corresponding expressions in the far field, $r\to \infty$, are much more complicated. 
Interestingly, their concrete form depends both on the value of $\lambda$ and the existence or not of an electric charge $Q$.
In the generic case, $Q \neq 0$ solutions, the asymptotic expansions are
\bea
w(r) = A e^{-r}  + \dots \, , ~~
V(r) = - \frac{Q}{r} + \dots \, , ~~
h(r) = 1 - \frac{4 \kappa_1 Q}{\lambda} \frac{1}{r^4} + \dots \, ,
\eea
for $\lambda \neq 0$,
 and
\bea
w(r)= A r^{-H_1} e^{-r}  + \dots \, ,~~
V(r)= - \frac{Q}{r} + \dots \, , ~~
h(r)= 1 + \frac{H_1}{r} + \dots \, , 
\eea
for $\lambda=0$, where $A$ and $H_1$ are constants.

However, for $Q=0$ one has to distinguish among several ranges for $\lambda$. For $\lambda>4$ we have
\bea 
w(r) = A e^{-r}  + \dots \, ,~~
V(r) = \frac{2\kappa_1 A^2}{r^2} e^{-2 r} + \dots \, , ~~
h(r) = 1 + \frac{2A^2}{4-\lambda} \frac{1}{r^2} e^{-2 r} + \dots \, , 
\eea
while for $0<\lambda<4$ we find 
\bea
w(r) = A e^{-r}  + \dots \, , ~~
V(r) =- \frac{4\kappa_1 H_1}{r^3} e^{-\sqrt{\lambda} r} + \dots \, ,~~ 
h(r)= 1 + \frac{H_1}{r} e^{-\sqrt{\lambda} r} + \dots \, . 
\eea
The corresponding expression for $Q=0,~\lambda=4$ reads
\bea 
w(r) =A e^{-r}  + \dots \, , ~~
V(r) = \frac{2\kappa_1 (A^2-H_0)}{r^2} e^{-2 r} + \dots \, , ~~
h(r) = 1 + H_0 e^{-2 r} + \dots \, , 
\eea 
while for $Q=0,~\lambda=0$
one finds
\bea
w(r)=A r^{-H_1} e^{-r}  + \dots \, , ~~
V(r) = - \frac{2\kappa_1H_1}{r^2} + \dots \, , ~~
h(r) = 1 + \frac{H_1}{r} + \dots \, . 
\eea

Interestingly, for these $Q=0$ solutions, we observe that the electric field decays exponentially, except for the $\lambda=0$ case where it exhibits a
dipole-like behaviour.

\begin{figure}[ht]
\begin{center}
\includegraphics[angle=-90,width=0.7\textwidth]{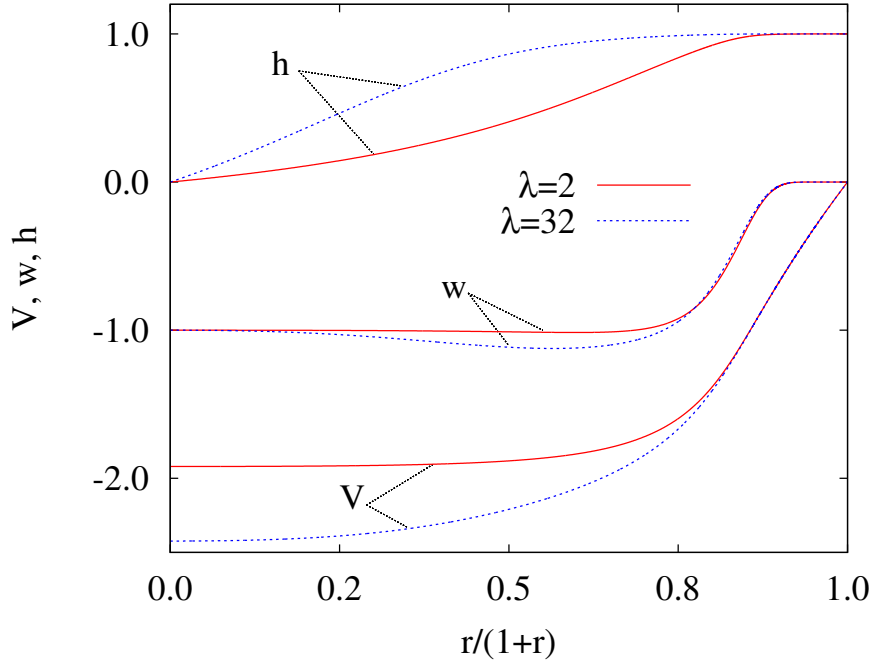}
\caption{Functions $w(r)$, $h(r)$, and $V(r)$ for two typical solutions with $\kappa_1=1$, $\kappa_2=3$ ($Q=6$), and $\lambda=2$ and $32$.}
\label{fig_functions}
\end{center}
\end{figure}

\section{Numerical results}

The system Eqs.~\re{eq_a}-\re{eq_h}
(with $V'$ given by Eq.~(\ref{firstintegral})) 
cannot be solved analytically and one has to resort to numerical methods to analyse its solutions. We have employed a
collocation method for boundary-value ordinary differential equations, equipped with an adaptive mesh selection procedure
\cite{COLSYS}. A compactified radial coordinate $x=r/(1+r)$ has been used. Typical mesh sizes include $10^3-10^4$ points.
The solutions have a relative accuracy of $10^{-10}$. 

In Fig.~\ref{fig_functions} we exhibit the functions $w$, $h$, and $V$ for two typical solutions with $\kappa_1=1$, $\kappa_2=3$ ($Q=6$), and
$\lambda=2,\ 32$. Since $Q$ does not vanish for these solutions, only function $w$ shows an exponential decay.

The effect of $Q$ on the functions is exhibited in Fig.~\ref{fig_function_b}, where the electric potential
 $V$ is plotted for solutions with $\kappa_2=-12$,
$\lambda=32$, and three values of $\kappa_1$: 1, 2, 3 ($Q=$-4, 0, 4, respectively). Only the solution with $Q=0$ gives rise to an exponential decay
for the electric potential (since $\lambda \neq 0$).

\begin{figure}[ht]
\begin{center}
\includegraphics[width=0.7\textwidth]{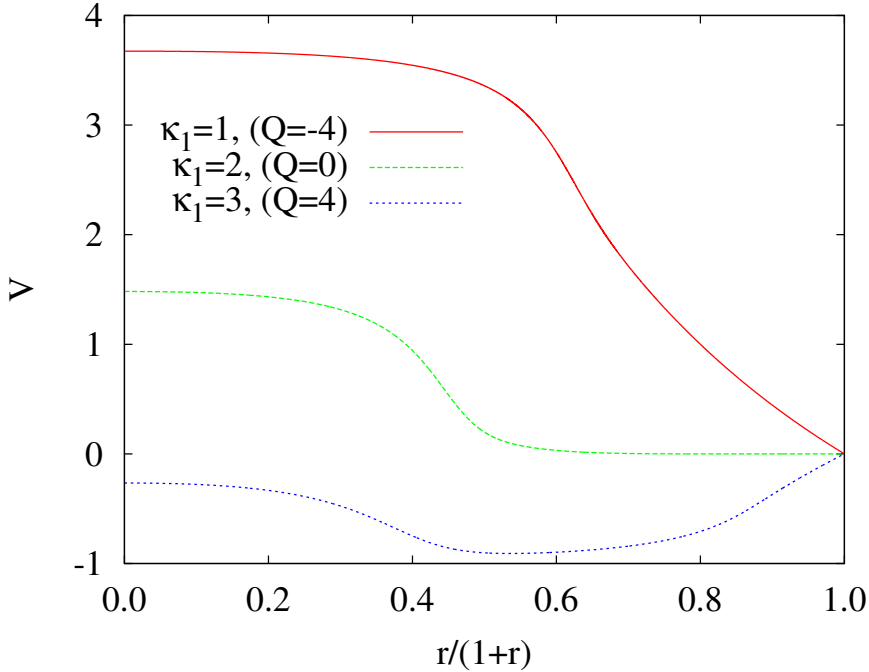}
\caption{The electric potential $V(r)$ for solutions with $\kappa_2=-12$, $\lambda=32$, and three values of $\kappa_1$: 1, 2, 3.}
\label{fig_function_b}
\end{center}
\end{figure}

Let us analyse the behaviour of the energy as a function of the parameters in the Lagrangian. For fixed $\lambda$, the energy depends on the CS coupling
constants $\kappa_1$ and $\kappa_2$. As a consequence of Eq.~\re{Q_expression}, the energy depends on $Q$ also. But we should emphasise again that $Q$ is
not a free parameter, but it is completely fixed once a concrete model is chosen (namely, once $\lambda$, $\kappa_1$, and $\kappa_2$ are chosen). However,
it is pertinent to ask what models produce the configuration with the lowest energy. If one sets any of the two CS coupling constants to be nonvanishing,
the configuration with the lowest energy does not correspond to the electrically uncharged one. We show an example of this in Fig.~\ref{fig_E_vs_Q}. Here
the energy of the solutions with $\kappa_2=-12$  and  $\lambda=32$ is plotted as a function of $Q$ (or equivalently, $\kappa_1$). It is seen that the
minimal energy occurs for $Q\neq0$.   

\begin{figure}[ht]
\begin{center}
\includegraphics[width=0.7\textwidth]{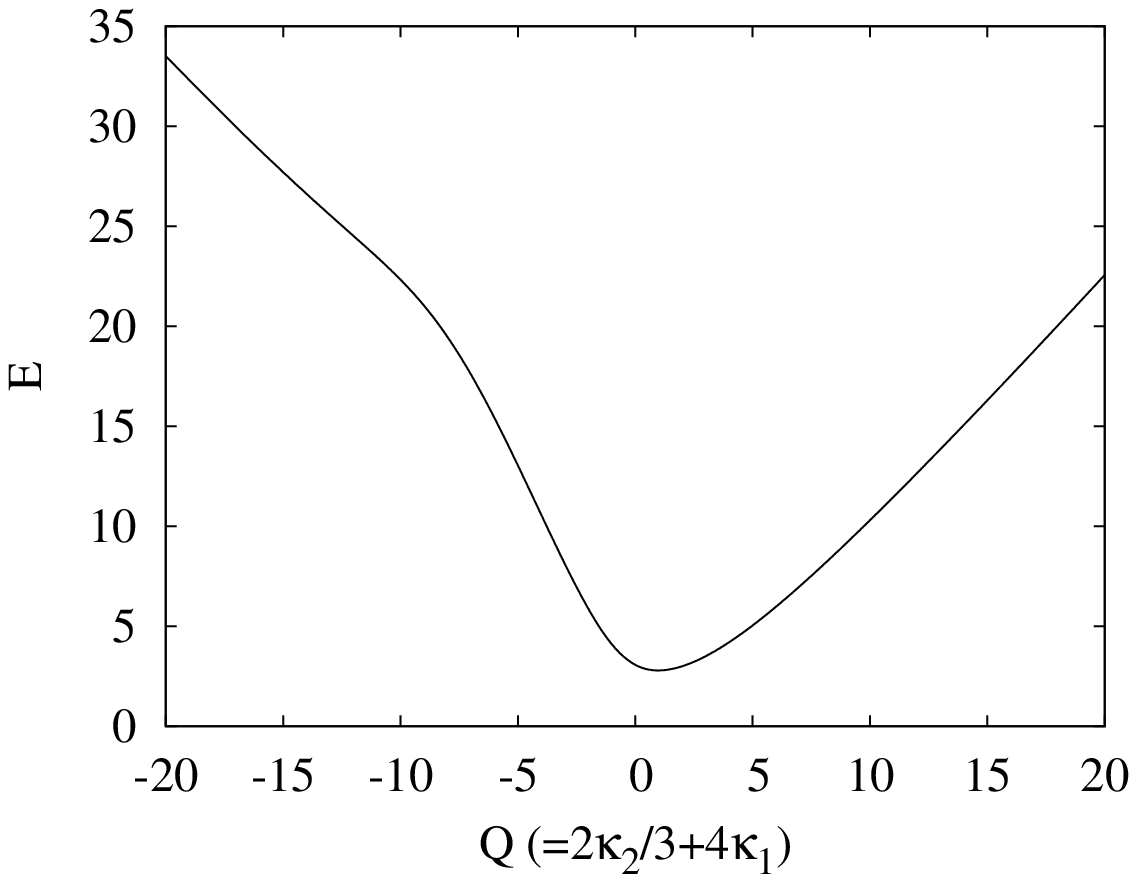}
\caption{Energy $E$  versus $Q$ for solutions with $\kappa_2=-12$  and  $\lambda=32$.}
\label{fig_E_vs_Q}
\end{center}
\end{figure}

One might be tempted to state that there might be a dyon whose energy was the lowest energy in this family of models. However, that does not seem to be
the case. We have explored large regions in the parameter space ($\kappa_1$, $\kappa_2$) for several values of $\lambda$ and the absolute minimal value
of the energy is always found to be that of the purely magnetic monopole, $i.e.$, $\kappa_1=\kappa_2=0$. We illustrate this fact in
Fig.~\ref{fig_E_vs_kappa_1_projected}. There, a 3D plot of $E$ versus $\kappa_1$ and $\kappa_2$ has been projected onto a $\kappa_2={\rm constant}$ plane
for $\lambda=32$ solutions. Energies of solutions with the same value of $\kappa_1$ correspond to points along a vertical line. We have highlighted the
energies of the $Q=0$ solutions (blue points). Clearly the absolute minimum occurs for $Q=0$ and $\kappa_1=0$ (which means $\kappa_2=0$). 

\begin{figure}[ht]
\begin{center}
\includegraphics[width=0.7\textwidth]{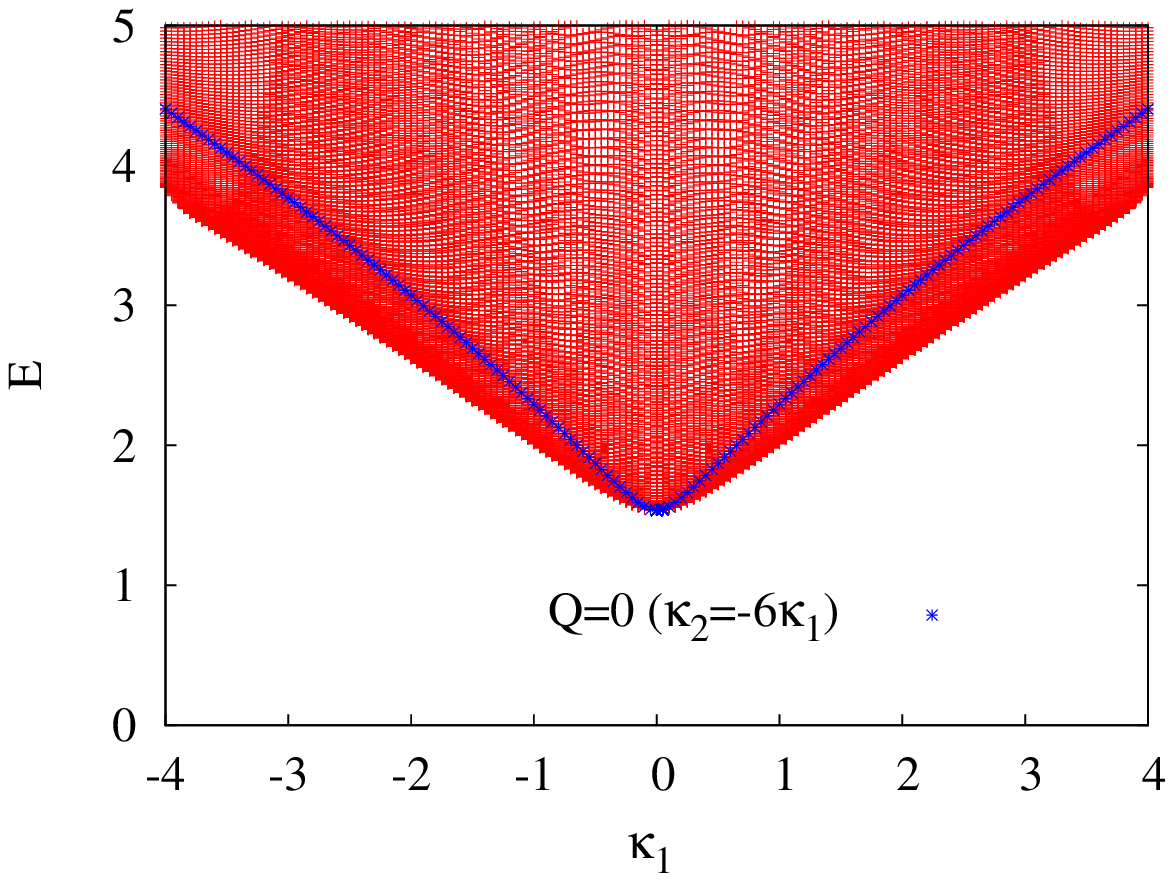}
\caption{Energy $E$  versus $\kappa_1$ for solutions with $\lambda=32$ for free $\kappa_2$ (projection onto a $\kappa_2={\rm constant}$ plane). Energies
of the $Q=0$ solutions are highlighted as blue star points.}
\label{fig_E_vs_kappa_1_projected}
\end{center}
\end{figure}

Finally, we will address the asymptotic behaviour of the energy as a function of $\lambda$. Here we observe that the effects of the two types of CS terms
are very different. In Fig.~\ref{fig_E_vs_lambda_asymptotics} we present the energy $E$ versus the Higgs potential coupling constant $\lambda$ in a
logarithmic scale for several values of $\kappa_1$ and $\kappa_2$. When $\kappa_2$ does not vanish, the energy diverges with $\lambda$. However, if
$\kappa_2=0$, the energy tends to a constant value (which depends on $\kappa_1$) as $\lambda$ tends to infinity, the same as for the usual, $SU(2)$
magnetic monopoles. The expression we found numerically for the asymptotic behaviour of the energy is
\be
E = c_1 \lambda^{1/4} + c_2 + c_3 \lambda^{-1/2} + \dots \, , \label{E_asymptotic}
\ee
where $c_1$, $c_2$, and $c_3$ are constants that depend on $\kappa_1$ and $\kappa_2$. $c_1$ vanishes for $\kappa_2=0$.

\begin{figure}[ht]
\begin{center}
\includegraphics[width=0.7\textwidth]{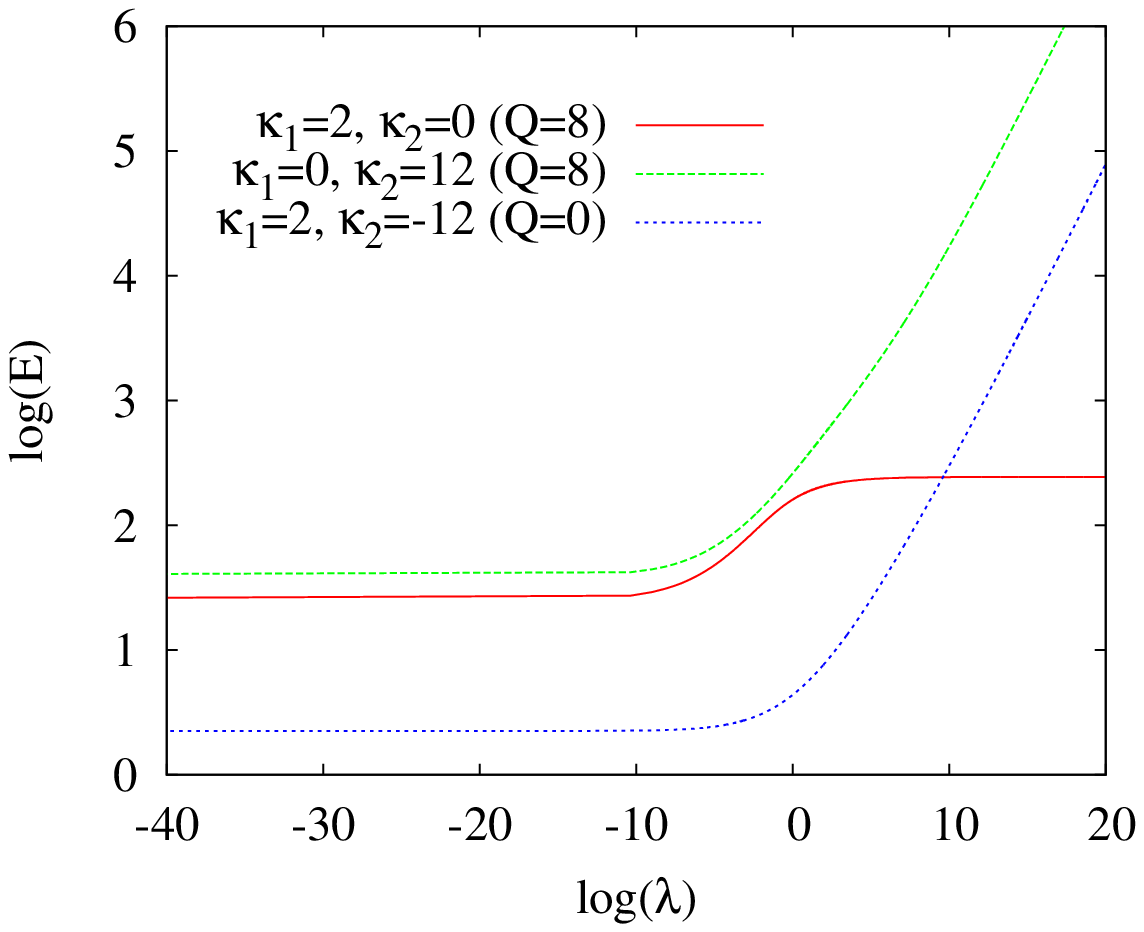}
\caption{Energy $E$ versus $\lambda$ for several values of $\kappa_1$ and $\kappa_2$.}
\label{fig_E_vs_lambda_asymptotics}
\end{center}
\end{figure}

\section{Further remarks}
The main purpose of this work was to provide an explicit construction
of spherically symmetric solitons of an $SO(5)$ Chern-Simons--Yang-Mills-Higgs theory in $3+1$ dimensional spacetime. To our knowledge,
no Chern-Simons solitons in even dimensional spacetime have appeared in the literature. The CS densities employed in this paper
are two of the $3+1$
dimensional ones introduced in \cite{Tchrakian:2010ar,Radu:2011zy}, which can be defined in spacetimes of all dimensions. In $D+1$ dimensions, these
are defined in terms of $SO(D+2)$ Yang-Mills field and a Higgs field taking its values in the orthogonal complement of $SO(D+2)$ in $SO(D+3)$. In any
given dimension there is an infinite tower of such densities, and here for the case $D=3$, we have considered the first two, Eqs.~\re{CS1} and \re{CS2},
in this tower~\footnote{These two CS densities are those extracted from the dimensional descent down to $3$ dimensions
of the $3$rd and $4$th Chern-Pontryagin densities in $6$ and $8$ dimensions, respectively. Higher order CS densities result from descents of
$n$-th CP densities in $n$ dimensions.}.

The solitons presented here differ from Julia-Zee~\cite{Julia:1975ff} dyons
in that the presence of the electric component of the gauge potential is a
result of the presence of the Chern-Simons density in the Lagrangian,
unlike in the case of the former~\cite{Julia:1975ff}. In the latter (JZ)
case, both the Higgs field (of the monopole) and the 'electric' component
of the gauge connection $A_0$ take their values in the algebra of $SO(3)$.
Here by contrast $A_0$ takes its values in the orthogonal complement of
$SO(3)$ in $SO(5)$, and the Higgs field takes its values in the orthogonal
complement of $SO(5)$ in $SO(6)$, $i.e.$ it is a $5$-component isovector.
They are CS dyons in the spirit of those appearing in \cite{Hong:1990yh,Jackiw:1990aw,NavarroLerida:2008uj}
in $2+1$ and \cite{Brihaye:2010wp} in $4+1$ dimensional spacetimes, respectively. The electric and magnetic fluxes here are defined by Eqs.~\re{elec} and
\re{mag}. Unlike for JZ dyons~\footnote{Non-Abelian JZ type dyons can be defined is all
spacetime dimensions~\cite{Tchrakian:2010ar} $D+1$, with $D\ge 3$.}, the global magnetic charge of our CS dyons is not the flux of a topological charge
density.

One should also point out that the various terms in the Lagrangian density, Eq.~\re{CSYMH}, have quite different dimensions. While such a
situation is not very unusual, it might nonetheless be that the qualitative features of the solutions may be different if all terms in the model had the
same dimensions. In the present context, the Yang-Mills-Higgs (YMH) model matching the dimensions of the CS density Eq.~\re{CS1} would be the sum of the
$p=1$ and the $p=2$ YMH model defined in Section {\bf 7.3} of \cite{Tchrakian:2010ar}, while the YMH model matching the dimensions of the CS density
Eq.~\re{CS2} would be the $p=2$ YMH model defined in Section {\bf 7.9} there~\cite{Tchrakian:2010ar}.

As avenues for future research, we mention first that it would be interesting to construct the generalizations of the solutions in this work
beyond the particular truncation Eq.~(\ref{truncation}),
$i.e.$ with full $SO(5)$ gauge potentials. 
However, so far we have encountered numerical difficulties in constructing the most general
solutions within the Ansatz (Eqs.~\re{higgs}-\re{aip}).
Another interesting direction 
would be to study the effects of gravity.
The study of gravitating monopoles and dyons  has started immediately after the discovery of these solitons
\cite{VanNieuwenhuizen:1975tc},
and has become a fruitful field of research (see \cite{Volkov:1998cc} for a review).
Moreover,
as suggested by a number of other gravitating models with non-Abelian Chern-Simons terms 
(see $e.g.$ 
\cite{Brihaye:2009cc},
\cite{Brihaye:2010wp},
\cite{Brihaye:2011nr}),
new features may occur in that case.

The inclusion of gravity effects can be approached in the standard way, by supplementing the 
Lagrangian Eq.~(\ref{CSYMH}) with an Einstein term and solving the corresponding field equations.
On general grounds,
one expects the existence of 
gravitating generalizations of the flat space solutions, at least for small
values of Newton's constant\footnote{One can show that the gravitating solutions
still possess a first integral similar to Eq.~(\ref{firstintegral}) which fixes their electric field.}. 
Moreover, the regular center can be replaced by a black hole with a (small enough) radius.
These solutions can be constructed by using similar techniques as those employed in Section 5.

However, we have found more interesting to consider gravitating generalization of the solutions with an extra
dilaton field coupled with both YM and Higgs sectors,
as described by the action  (here we consider the first CS model only)
\begin{eqnarray}
\label{action-grav}
S=\frac{1}{4\pi G}
\int d^4 x \sqrt{-g}
\left[
\frac{1}{4}R
-\frac{1}{2}\partial_\mu\varphi \partial^\mu\varphi
-\frac{e^{c_1 \varphi}}{4}\mbox{Tr}(F_{\mu\nu}F^{\mu\nu})
-\frac{e^{c_2 \varphi}}{2}\mbox{Tr}(D_{\mu }\Phi D^{\mu }\Phi)
-\frac{i \kappa_1}{\sqrt{-g}} \epsilon^{\mu \nu\rho \sigma} \mbox{Tr} (F_{\mu\nu}F_{\rho \sigma})
\right ]~.
\end{eqnarray}
Then, following the prescription in \cite{Gibbons:1993xt}, one can show that for 
the $SO(3)\times{SO(2)}$  subsystem with
\begin{eqnarray}
\label{rel1}
c_1=-\frac{1}{2}c_2=\frac{2}{\sqrt{3}},~~\kappa_1=\frac{1}{2\sqrt{3}}~,
\end{eqnarray}
the flat spacetime $SO(3)$ BPS monopoles remain a solution of the theory.
Moreover, all
other fields have simple closed form expressions in this case.
Restricting again to the spherically symmetric case, the corresponding solution has a line element
\begin{eqnarray}
\label{metric}
ds^2=U(r)^{3/2}\left(dr^2+r^2(d\theta^2 +\sin^2\theta d\phi^2) \right)-\frac{dt^2}{U(r)^{3/2}}~,
\end{eqnarray}
with
\begin{eqnarray}
\label{relU}
U(r)=1+\frac{C}{r}-\frac{2}{3r^2}+\frac{4\coth r}{3r}-\frac{2}{3\sinh^2r}~,
\end{eqnarray}
where $C$ is an arbitrary integration constant.
The matter fields have the following expression
\begin{eqnarray}
\label{dilaton}
\varphi(r)=\frac{\sqrt{3}}{4}\log U(r),~~V(r)=\frac{\sqrt{3}}{2 U(r)},~~
w(r)=\frac{r}{\sinh r},~~h(r)=\coth r-\frac{1}{r}~.
\end{eqnarray}
For $C>0$, 
this describes an extremal black hole solution with non-Abelian hair,
with an horizon located at $r=0$.
However, similar to the Einstein-Maxwell-dilaton case, 
$r=0$ is a naked singularity, since 
the Kretschmann scalar diverges as $1/r$ as $r\to 0$.
Taking $C<0$ leads again to a singular configuration
(the singularity occurs this time for some $r_0>0$).

Thus the only interesing case corresponds to $C=0$.
The basic properties of this configuration are discussed already in \cite{Gibbons:1993xt}
(although the explicit relation Eq.~(\ref{relU}) is not given there).
As proved in \cite{Gibbons:1993xt},
this solution describes a globally regular gravitating soliton,
with both electric and magnetic charges,
whose mass equals the magnetic charge.

Let us close by remarking that the $SO(3)\times{SO(2)}$ Ansatz in Section 3
can be generalized by including a winding number in the $SO(3)$ sector.
This would lead to axially symmetric magnetic
monopoles endowed,  via the CS term, with an extra electric charge.
A similar construction to that described above would lead to closed form
gravitating solutions whose geometry is static and axially symmetric.


\medskip
\medskip

\noindent
{\bf\large Acknowledgements}
D.H.T. is grateful to Professor Hermann Nicolai for his hospitality at the Albert-Einstein-Institute, Golm,
(Max-Planck-Institut, Potsdam) where this work was started. E.R. gratefully acknowledges support from the FCT-IF programme. This work was carried
out in the framework of the Spanish Education and Science Ministry under Project No. FIS2011-28013.


\begin{small}

\end{small}

\end{document}